\documentclass{emulateapj}
\usepackage{epstopdf}

\newcommand {\aplt} {\ {\raise-.5ex\hbox{$\buildrel<\over\sim$}}\ }

\slugcomment{Accepted for Publication in ApJ Letters}
\shorttitle{Late-Time Observations of PS1-10jh}
\shortauthors{Gezari et al.}

\begin{document}

\title{PS1-10jh Continues to Follow the Fallback Accretion Rate of a Tidally Disrupted Star}

\author{S. Gezari\altaffilmark{1}, R. Chornock\altaffilmark{2}, A. Lawrence\altaffilmark{3}, A. Rest\altaffilmark{4}, D. O. Jones\altaffilmark{5}, E. Berger\altaffilmark{6}, P. M. Challis\altaffilmark{6}, \& G. Narayan\altaffilmark{7}
}
\altaffiltext{1}{Department of Astronomy, University of Maryland, Stadium Drive, College Park, MD 20742-2421, USA \email{suvi@astro.umd.edu}}
\altaffiltext{2}{Astrophysical Institute, Department of Physics and Astronomy, 251B Clippinger Lab, Ohio University Athens, OH 45701, USA}
\altaffiltext{3}{Institute for Astronomy, University of Edinburgh Scottish Universities Physics Alliance, Royal Observatory, Blackford Hill, Edinburgh EH9 3HJ, UK}
\altaffiltext{4}{Space Telescope Science Institute, 3700 San Martin Drive, Baltimore, MD 21218, USA}
\altaffiltext{5}{Department of Physics and Astronomy, Johns Hopkins University, 3400 North Charles Street, Baltimore, MD  21218, USA}
\altaffiltext{6}{Harvard-Smithsonian Center for Astrophysics, 60 Garden Street, Cambridge, MA  02138, USA}
\altaffiltext{7}{NOAO, 950 North Cherry Avenue, Tucson, AZ 85719, USA}

\begin{abstract}
We present late-time observations of the tidal disruption event candidate PS1-10jh.  UV and optical imaging with HST/WFC3 localize the transient to be coincident with the host galaxy nucleus to an accuracy of 0.023 arcsec, corresponding to 66 pc.  The UV flux in the F225W filter, measured 3.35 rest-frame years after the peak of the nuclear flare, is consistent with a decline that continues to follow a t$^{-5/3}$ power-law with no spectral evolution.  Late epochs of optical spectroscopy obtained with MMT $\sim$ 2 and 4 years after the peak, enable a clean subtraction of the host galaxy from the early spectra, revealing broad helium emission lines on top of a hot continuum, and placing stringent upper limits on the presence of hydrogen line emission.  We do not measure Balmer H$\delta$ absorption in the host galaxy strong enough to be indicative of a rare, post-starburst "E+A" galaxy as reported by \citet{Arcavi2014}.  The light curve of PS1-10jh over a baseline of 3.5 yr is best modeled by fallback accretion of a tidally disrupted star.  Its strong broad helium emission relative to hydrogen (He II $\lambda 4686$/H$\alpha > 5$) could be indicative of either the hydrogen-poor chemical composition of the disrupted star, or certain conditions in the tidal debris of a solar-composition star in the presence of an optically-thick, extended reprocessing envelope.  \end{abstract}


\section{Introduction}

The tidal disruption of stars was first proposed by theorists as an inevitable dynamical consequence of a supermassive black hole in the nucleus in a galaxy \citep{Lidskii1979, Rees1988}.  While the first tidal disruption event (TDE) candidates were discovered in archival studies of X-ray surveys \citep[see review by][]{Komossa2002}, the detection of TDE candidates in high-cadence optical surveys such as CFHT SNLS \citep{Gezari2008}, SDSS Stripe 82 \citep{vanVelzen2011}, Pan-STARRS1 \citep{Gezari2012, Chornock2014}, PTF \citep{Arcavi2014}, and ASASSN \citep{Holoien2014, Holoien2015}, have yielded light curves with excellent temporal sampling, such that predictions for the time-dependent radiation produced from the accretion flow in a TDE are being confronted in detail for the first time.  

While it is encouraging that the optical TDE candidates have roughly the right luminosities, rates, and timescales predicted by basic TDE theory,   there are some major disagreements between the theory and observations that remain to be resolved.  i) Predictions for the evolution of the hot, thermal continuum in a newly-formed debris disk in a TDE \citep{Ulmer1999, Lodato2011, Guillochon2014} appear to be in conflict with the constant, relatively cooler ($T_{\rm eff} \sim 1-3 \times 10^{4}$ K) temperatures observed in the optical TDE candidates.  ii) The optical light curve shapes seem to agree well with the rise time and canonical $t^{-5/3}$ power-law decay expected from the fallback of debris from a tidally disrupted star, however, this is surprising considering the potentially long circularization timescales, especially in low-mass black holes \citep{Shiokawa2015, Hayasaki2015, Guillochon2015}, and the expected band effects in the optical from cooling of the emission with time \citep{Strubbe2009, Lodato2011, Guillochon2014}.  iii) The total energies emitted are much smaller than expected for the tidal disruption and accretion of a star, and may suggest that either partial disruptions are common \citep{MacLeod2013, Chornock2014}, or the fraction of debris that is available for accretion on to the black hole is much smaller than the fraction of initially bound debris \citep{Ayal2000, Metzger2015} 

Solutions for these discrepancies have been proposed in the literature, including temperature regulation via winds \citep{Miller2015}, invoking a reprocessing layer \citep{Loeb1997, Guillochon2014}, a radiation-dominated nebula \citep{Metzger2015, Strubbe2015}, or attributing the continuum emission directly to the circularization process \citep{Piran2015}.

Here we present late-time observations of the TDE candidate PS1-10jh \citep[][hereafter G12]{Gezari2012}.  This TDE candidate is of particular importance for two reasons.  First, it has by far the best sampled UV/optical light curve of a TDE candidate to date.  
Second, its optical spectrum at peak was dominated by broad He II
emission lines, strongly suggesting that the photoionized
gas did not originate from the quiescent interstellar medium in the vicinity of the SMBH, or in a pre-existing accretion disk, 
but rather from the tidal debris of the disrupted star itself.
Our aim in this study was to measure the UV flux of the event at late times, in order to localize the transient relative to its host galaxy nucleus with the high spatial resolution of HST, and compare its continued decay with models for the evolution of the accretion flow over time.  In addition, we obtained a deeper late-time spectrum of the host galaxy, in order to construct a more reliable host-subtracted spectrum of the nuclear transient. 

The paper is organized as follows.  We present our HST observations in \S \ref{sec:hst}, and in \S \ref{sec:analysis} we use the late-time UV emission measured by HST to pinpoint its origin to the nucleus of the host galaxy and its central SMBH, and probe the evolution of its light curve over a baseline of over 3 years after its peak.  In \S \ref{sec:spec} we use late-time optical spectroscopy to perform a clean subtraction of the host galaxy contribution, and isolate the non-stellar continuum and emission lines powered by PS1-10jh.  In \S \ref{sec:disc} we discuss the implications of these observations with respect to the geometry of the accretion flow, the evolution of its effective temperature with time, and the chemical composition of the star disrupted.

\section{HST Observations} \label{sec:hst}
We obtained HST observations of PS1-10jh on UT 2014 June 13 with WFC3 UVIS in the F225W and F625W filters for 3.791 ksec and 5.014 ksec, respectively, with the goal of localizing the transient with respect to its host galaxy nucleus, and measuring its late-time UV flux.  The source is clearly detected as a point source in the F225W filter, with no evidence of extended host emission, consistent with its deep \textsl{GALEX} pre-event imaging upper limit of $> 25.6$ mag (G12).  The optical image is dominated by the host galaxy, and we use it to determine the precise position of the host galaxy centroid.

\begin{figure}
\plotone{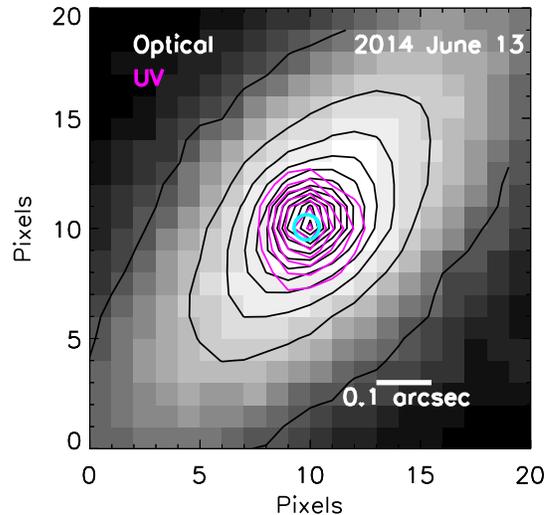}
\caption{
HST WFC3/UVIS image of PS1-10jh taken on 2014 June 13 in the F225W (UV) and F625W (optical) filters.  Grey scale and black contours show optical image dominated by the host galaxy, and magenta contours show the UV point source associated with the fading transient, 3.35 rest-frame years after its peak.   1 $\sigma$ error circle for F225W source centroid (0.57 pixels = 0$\farcs$023) plotted with a cyan circle.  The position of the UV source is coincident with the nucleus of the host galaxy within 1 $\sigma$. \label{fig:centroid} }
\end{figure}

\section{Analysis} \label{sec:analysis}
\subsection{Position Relative to Host Nucleus}

We register the UV (F225W) and optical (F625W) images to each other using 7 reference stars in the field of view.  We use the IRAF routines {\tt geomap} and {\tt geotran} to calculate and apply a geometrical transform of the UV image to the optical image.  After this registration, we find that the source is coincident with the optical host galaxy centroid within $1\sigma$ = 0.57 pixels = $0\farcs 023$ (see Figure \ref{fig:centroid}).   The host galaxy is well resolved in the optical image, and is fitted with elliptical isophotes with a de Vaucouleurs surface brightness profile, and an effective (half-light) radius, $r_e = 0\farcs26$.  

\subsection{Late-Time UV Flux}

We perform aperture photometry on the UV image using the IDL routine {\tt APER}, using a $0\farcs2$ radius aperture, and correcting for a fraction of 0.777 of the total energy enclosed at the filter effective wavelength of 235.9 nm, and find $m = 23.69 \pm 0.05$ mag in the AB system, and $NUV = 23.70 \pm 0.12$ mag in the \textsl{GALEX} AB system, measured from comparison to the \textsl{GALEX} AB magnitudes measured with a 6 arcsec radius aperture for 4 isolated point sources measured from a 28 ksec coadd (5$\sigma$ detection limit of $m_{\rm lim} = 24.7$ mag) constructed from observations taken during the \textsl{GALEX} Time Domain Survey \citep{Gezari2013}.

\begin{figure}
\plotone{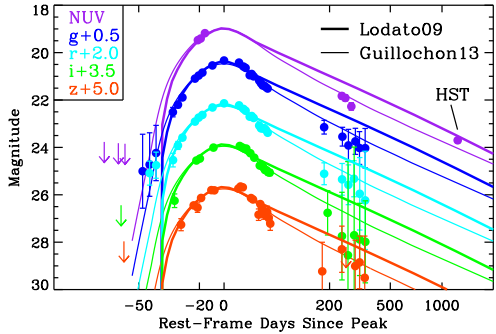}
\caption{UV/optical light curve of tidal disruption event candidate PS1-10jh as measured by \textsl{GALEX} in the $NUV$ and PS1 in the $g,r,i,z$ bands, along with the late-time UV point measured by HST/WFC3 UVIS in the F225W filter 1250 rest-frame days since the peak of the flare.  Models for the fallback rate of a tidally disrupted star from \citet{Lodato2009} are plotted with thick lines and from \citet{Guillochon2013} are plotted with thin lines.  The late-time UV flux measured by HST is consistent with a continued decline close to the canonical $t^{-5/3}$ power-law expected from the fallback rate for a tidally disrupted star.
\label{fig:lc}
}
\end{figure}

We show the UV/optical light curve of PS1-10jh from \textsl{GALEX} and PS1 in Figure \ref{fig:lc}, with the late-time HST epoch in the NUV (F225W) filter at MJD 56821.434 (UT 2016 June 13.497), or 3.35 rest-frame years from the observed peak of the light curve.  We have improved the PS1 image differencing photometry from G12 using the methods of \citet{Rest2014}.  Making the (not necessarily correct) assumption that the emerging flux is directly proportional to the fallback rate, we fit the UV/optical light curve to models for the fallback rate, with different values for the polytropic exponent ($\gamma$) of the disrupted star, and the impact parameter of its orbit defined as $\beta \equiv r_t/r_p$.   We show a fit to a model for a star with $\gamma = 5/3$ and $\beta = 1$ model from \citet{Lodato2009} presented in G12, and $\beta=  1.8$ and $\gamma = 4/3$ model following the methodology of \citet{Guillochon2013} (J. Guillochon, private communication).  The time delay between the time of disruption ($t_D$) in the models, and the peak of the fallback rate ($t_0$), are $\Delta t = t_0-t_D = 76$ and 84 rest-frame days, respectively.  The corresponding black hole masses for each model are $M_{\rm BH} \sim 2$ and $9 \times 10^{6} M_\odot r_\star^{-3} m_\star^2$, where $r_\star = R_\star/R_\odot$ and $m_\star=M_\star/M_\odot$.  Independent of these models at late times, but adopting their values of $t_D$ (which is tightly constrained from the rising portion of the light curve), the UV decline of PS1-10jh is fitted with a power-law index, $\propto (t-t_D)^{-n}$, with $n = 1.66 \pm 0.03$, in nice agreement with the analytical $t^{-5/3}$ power-law calculated for fallback  of tidally disrupted debris \citep{Rees1988, Phinney1989}.

The different black hole masses from the model fits are due to the different assumption in $\gamma$ as well as $\beta$.  Note, that these are fits to the fallback rate $\dot{M}$, and do not include radiative transfer or band effects.  For example, even if the bolometric luminosity is tracking $\dot{M}$, the flux in a given band will depend on the evolution of the spectral energy distribution (SED) with time.  However, given the stable UV/optical colors in PS1-10jh over a year after peak, there is little evidence of cooling or of a changing accretion efficiency with time.  Thus with a fixed SED shape, both $L_{\rm bol}$ and $L_\nu$ will follow the decline of $\dot{M}$.  

Remarkably, the late-time UV flux measured by HST continues to follow the decline expected from the fallback rate predicted by both models.  This persistent $t^{-5/3}$ power-law decline in the UV at late times again is evidence for both the lack of cooling (or heating) of the emission.  While for a fixed radius the temperature of thermal emission would be expected to cool with a declining luminosity, the temperature of the emission might actually increase at late times due to a shrinking photosphere from a radiatively-driven wind \citep{Lodato2011}.  A change in the SED temperature would cause the UV light curve to flatten or steepen, neither of which is observed.  Furthermore, we don't see evidence for viscous evolution, which is expected to occur on the timescale of tens to hundreds of years, and would result in a flatter power-law \citep{Cannizzo1990, Shen2014}.

The fact that the UV/optical emission so closely follows the fallback rate is in contradiction with debris disk models \citep{Ulmer1999, Lodato2011, Guillochon2013}, and requires a reprocessing layer or outflow with a regulated temperature \citep{Guillochon2013}, and may be the natural behavior of a radiation-dominated wind \citep{Miller2015} or expanding nebula \citep{Metzger2015}, albeit with an unphysical receding photosphere.   Alternatively, the optical emission may arise from the circularization process itself, and not accretion \citep{Piran2015}.  While the larger emission radius in this scenario would produce an effective temperature well matched to PS1-10jh, the shock dissipation mechanism powering the emission is not expected to dissipate energy as smoothly, and monotonically evolving as observed in PS1-10jh. 

\section{Late-Time Spectroscopy} \label{sec:spec}

We obtained two additional late-time epochs of spectroscopy with the 6.5m MMT Blue Channel spectrograph \citep{Schmidt1989} on UT 2012 September 16--17 ($2\times1800$ sec each night) and 2014 April 28 ($3\times1800$ sec).  We used a 1$\arcsec$ slit at the parallactic angle \citep{Filippenko1982} and the 300 lines/mm grating, for a spectral resolution of $\sim$ 5.5 \AA. This was the same instrumental setup used for the Day $-22$ and $+254$ observations in G12.  When we subtract the 2012 and 2014 spectra from each other, we find no significant differences within the errors.  At these epochs the optical emission from PS1-10jh has faded by factors $> 25$, and thus is no longer detectable above the host galaxy continuum.  Thus we coadd these late-time spectra to create a host galaxy template.  Note that we measure a Balmer H$\delta$ absorption equivalent-width (EW) in the host galaxy spectrum of 2.3 $\pm 0.3$ \AA, which does not classify the galaxy as a rare post-starburst E+A galaxy \citep{Goto2003}, as claimed by \citet{Arcavi2014}.   

\begin{figure*}
\plottwo{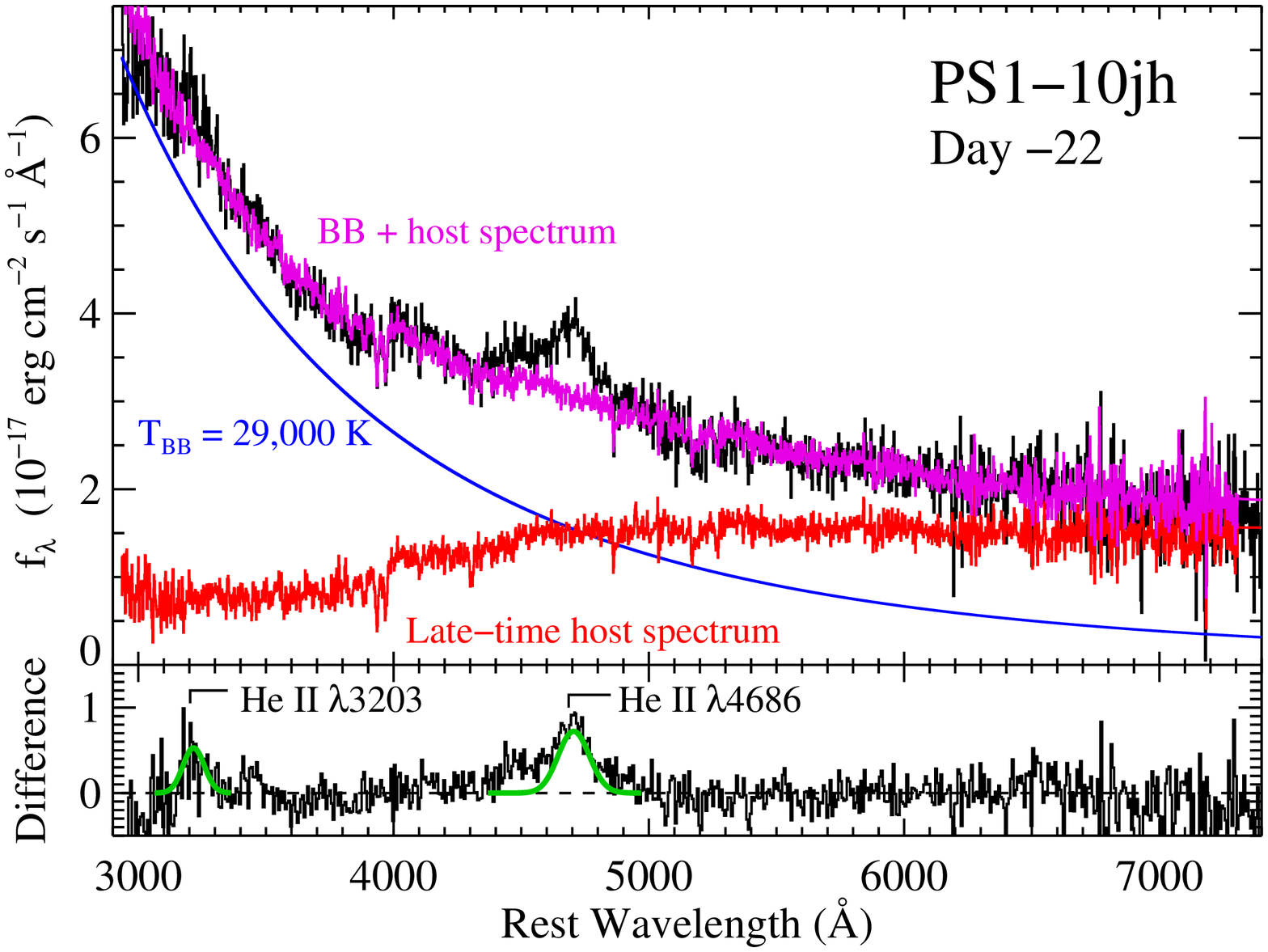}{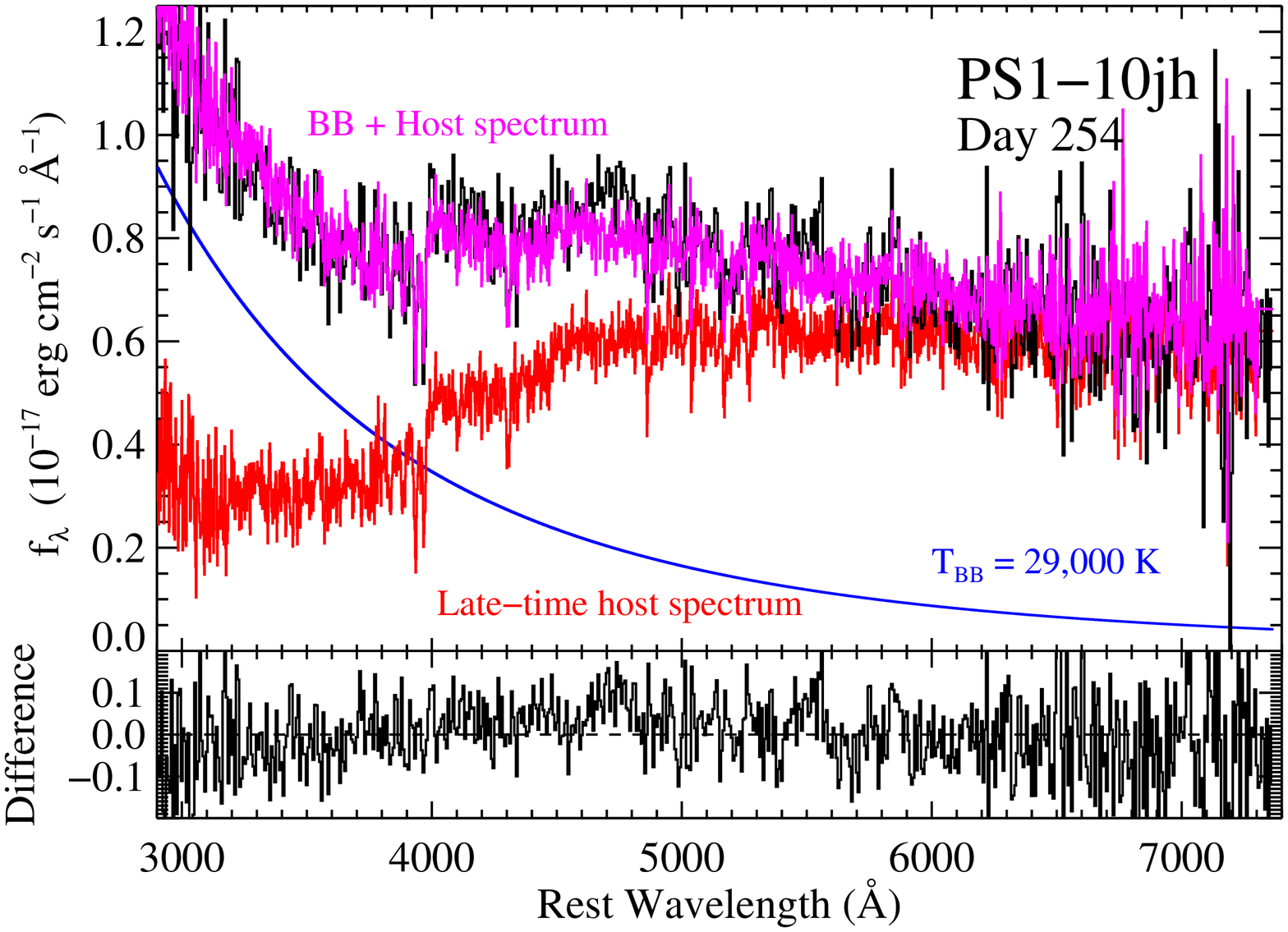}
\caption{Two epochs of MMT spectra of PS1-10jh obtained $-22$ and $+254$ days from the peak of the flare.  Spectrum is plotted in black, hot-blackbody continuum plus host galaxy spectrum plotted in magenta, hot blackbody component plotted in blue, and difference spectrum plotted in lower panel.  Gaussian fits to the He II $\lambda$ 3203 and 4686 lines plotted in green.
\label{fig:spec}
}
\end{figure*}

We subtract the host galaxy plus a hot blackbody ($T_{\rm bb} = 2.9 \times 10^{4}$ K) at $z=0.1696$ from the earlier spectral epochs presented in G12, in order to remove the continuum from PS1-10jh and its host, and isolate its emission line features.  The resulting difference spectrum in Figure \ref{fig:spec} is very similar to that presented in G12, in which we used a stellar-synthesis galaxy template subtraction, although there is some loss of S/N due to the noise in the true host spectrum.  We detect broad He II emission lines at $\lambda 4686$, and $\lambda 3203$, including a "blue wing" to He II $\lambda$4686.    This feature was also present in the original analysis, although we did not discuss it, since there was the possibility that this was an artifact from a template mismatch.   Now that it has been confirmed to be real, it could result from an outflow or other velocity structure in the emission line, or the presence of additional lines, such as the CIII/NIII "Wolf-Rayet" blends seen in some supernovae (SNe) when the the temperature is high \citep[e.g.,][]{Niemela1985, Leonard2000}.  However, the second component can be modeled as a second Gaussian with a centroid near 4470 \AA, which is too blue for the WR blends at zero velocity.  Note that TDE candidate ASASSN-14li \citep{Holoien2015} also very clearly has velocity structure in its He II 4686 line, although it is double-peaked at a different velocity, so it is not the same line blended in the two objects.  

\begin{figure}
\plotone{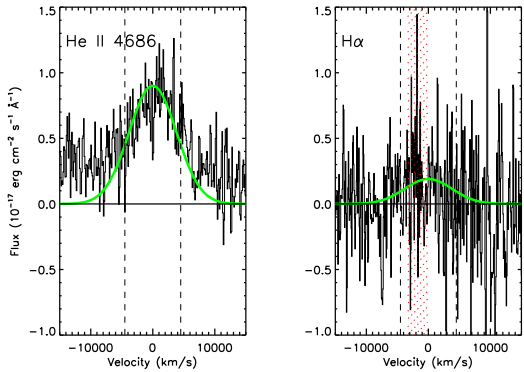}
\caption{Zoom in of He II $\lambda 4686$ and H$\alpha$ lines.  Dashed lines show the 9,000 km/s FWHM of the He II line.  Adding up the flux in this velocity range yields He II $\lambda$ 4868/H$\alpha$ = 4.7.  Gaussians with this ratio (and a FWHM equal to the He II line) are plotted in green.  Shaded red region shows the velocity range of the telluric A-band feature, where systematic errors in the telluric correction may be present.
\label{fig:ha}
}
\end{figure}

The most striking aspect of the spectrum is still the lack of obvious hydrogen emission lines.  Figure \ref{fig:ha} shows a zoom in of the spectrum at the rest-frame velocity of He II $\lambda 4686$ and H$\alpha$.  Dashed lines show the 9,000 km s$^{-1}$ FWHM of the Gaussian fit to the He II $\lambda 4686$ line measured in G12.  If we integrate the flux in this velocity range for both lines, we measure a ratio of  He II $\lambda 4686$/H$\alpha = 4.7 \pm 1.0$.  In Figure \ref{fig:ha} we plot Gaussians with a FWHM= 9,000 km s$^{-1}$ that have this flux ratio.   Note that H$\alpha$ at this redshift is in the wing of the telluric A band, which corresponds to wavelengths in the rest-frame of PS1-10jh of $\sim 6490-6560$ \AA.  Furthermore, He II has a line at 6560 \AA (the Pickering series $n=6\rightarrow 4$), in wavelength coincidence with H$\alpha$, but it is expected to be only $\sim 0.14$ times the strength of He II 4686 for photoionized gas \citep{Osterbrock1989}.
Despite some of these systematic uncertainties, we do not find any significant residual emission at H$\alpha$.  \citet{Gaskell2014} reported residual emission at H$\alpha$ corresponding to He II $\lambda 4686$/H$\alpha$ = 3.7 based on a smoothed version of the model galaxy template subtraction we performed in G12.  However, our late-time spectroscopy now enable us to perform a direct host galaxy subtraction that is more accurate, and is still consistent with the He II $\lambda 4686$/H$\alpha > 5$ ratio reported in G12.  
H$\beta$ is more difficult to measure an upper limit for, since it is located within the red wing of the much stronger broad He II $\lambda 4686$ line.  

\section{Discussion \& Conclusions} \label{sec:disc}

The nuclear position of PS1-10jh in its host galaxy further strengthens its association with the host galaxy's central SMBH, and disfavors a SN origin, or a microlensing event by a star in an (unseen) foreground galaxy \citep{Lawrence2012}.   Similarly, the persistence of the UV emission on the timescale of years is inconsistent with any known SN behavior.  
The continued decline of the
UV light-curve at the rate expected for the fallback accretion of the tidally disrupted debris is incompatible with simple accretion-flow geometries, and requires an extended photosphere that evolves in such a way to keep the effective temperature fixed with time.

The strong detection of broad helium emission, and lack of hydrogen emission, is difficult to reproduce in standard conditions for photoionized gas.  In G12, we argue that the lack of hydrogen emission is a signature of a low hydrogen fraction ($X < 0.2$) for the disrupted star.  However, helium stars are extremely rare, and even tidally stripped evolved stars retain a large fraction of their H envelopes \citep{MacLeod2013}, and would require multiple orbital passages of the SMBH to completely unbind their H envelope \citep{Bogdanovic2014}.  Subsequent arguments have been made that avoid the problem of tidally disrupting a helium-rich star, by suppressing the hydrogen emission from a truncated broad-line region formed from the tidal debris of a main-sequence star \citep{Guillochon2014}, in particular, at certain densities where thermalization becomes important \citep{Gaskell2014}.  However, \citet{Strubbe2015} find that these conditions in solar-composition photoionized gas are only sufficient to suppress the H$\alpha$ emission enough to yield He II $\lambda 4686$/H$\alpha \sim 3$, or in the presence of velocity gradients in the gas, only up to He II $\lambda 4686$/H$\alpha  < 1$, well below the line ratio we measure for PS1-10jh of He II $\lambda 4686$/H$\alpha > 5$.  While recent calculations by \citet{Roth2015} find that certain conditions in an {\it optically-thick} extended reprocessing envelope in a TDE could reproduce a ratio of He II $\lambda 4686$/H$\alpha = 5$, or even higher.

While at least one other TDE candidate has a similar spectrum to PS1-10jh \citep{Arcavi2014}, there are also TDE candidates reported with both hydrogen and helium broad emission lines \citep{Arcavi2014, Holoien2014, Holoien2015}, as well as neither (only absorption) \citep{Chornock2014}.  Clearly, hydrogen emission is not suppressed in all TDE candidates.  While this may be a signature of the diversity in the hydrogen envelopes of disrupted stars, analogous to the hydrogen-free spectra of stripped core-collapse SNe, the radiative transfer effects in the complex photoionized debris stream or disk may also be a natural explanation, and must continue to be investigated in more detail before stronger conclusions can be made.  Detailed spectroscopic monitoring of future TDEs is critical for testing these models, directly probing the structure of the stellar debris in a TDE, and providing insights on the population of stars in the harsh environment in the vicinity of its central SMBH.

\acknowledgements

We thank the anonymous referee for their helpful comments.  S.G. was supported in part by NSF CAREER grant 1454816.  RC thanks the Aspen Center for Physics, which is supported by National Science Foundation grant PHY-1066293, for their hospitality while some of this work was performed.  Based on observations made with the NASA/ESA Hubble Space Telescope associated with program \# 13371.  Support for program \# 13371 was provided by NASA through a grant from the Space Telescope Science Institute, which is operated by the Association of Universities for Research in Astronomy, Inc., under NASA contract NAS 5-26555.  Some of the observations reported here were obtained at the MMT Observatory, a joint facility of the Smithsonian Institution and the University of Arizona.  


\clearpage

\end{document}